\documentclass{vgtc}                          

\graphicspath{{figures/}{pictures/}{images/}{./}} 
\usepackage{float}
\usepackage{times}                     

\usepackage{tabu}                      
\usepackage{booktabs}                  
\usepackage{lipsum}                    
\usepackage{mwe}                       
\usepackage{subcaption}
\usepackage{mathptmx}                  
\usepackage{amsmath}
\usepackage{enumitem}
\usepackage[most]{tcolorbox}

\definecolor{delcolor}{HTML}{8B0000}      
\definecolor{addcolor}{HTML}{00008B}      
\definecolor{revisecolor}{HTML}{CC6600}   

\onlineid{0}

\vgtccategory{Research}

\vgtcinsertpkg

\title{Do Visualization Literacy Assessment Tests Account for High-Level Visualization Comprehension?}

\author{Aaryani Chowdary Ambati\textsuperscript{*,\ensuremath{\dagger}}\\ %
        \scriptsize University of Oklahoma %
\and Tapendra Pandey\textsuperscript{*,\ensuremath{\dagger}}\\ %
        \scriptsize University of Oklahoma %
\and Arran Zeyu Wang\textsuperscript{\ensuremath{\ddagger}}\\ %
        \scriptsize UNC-Chapel Hill %
\and Ghulam Jilani Quadri\textsuperscript{\ensuremath{\dagger}}\\ %
        \scriptsize University of Oklahoma %
}

\teaser{
  \centering
  \includegraphics[width=\linewidth]{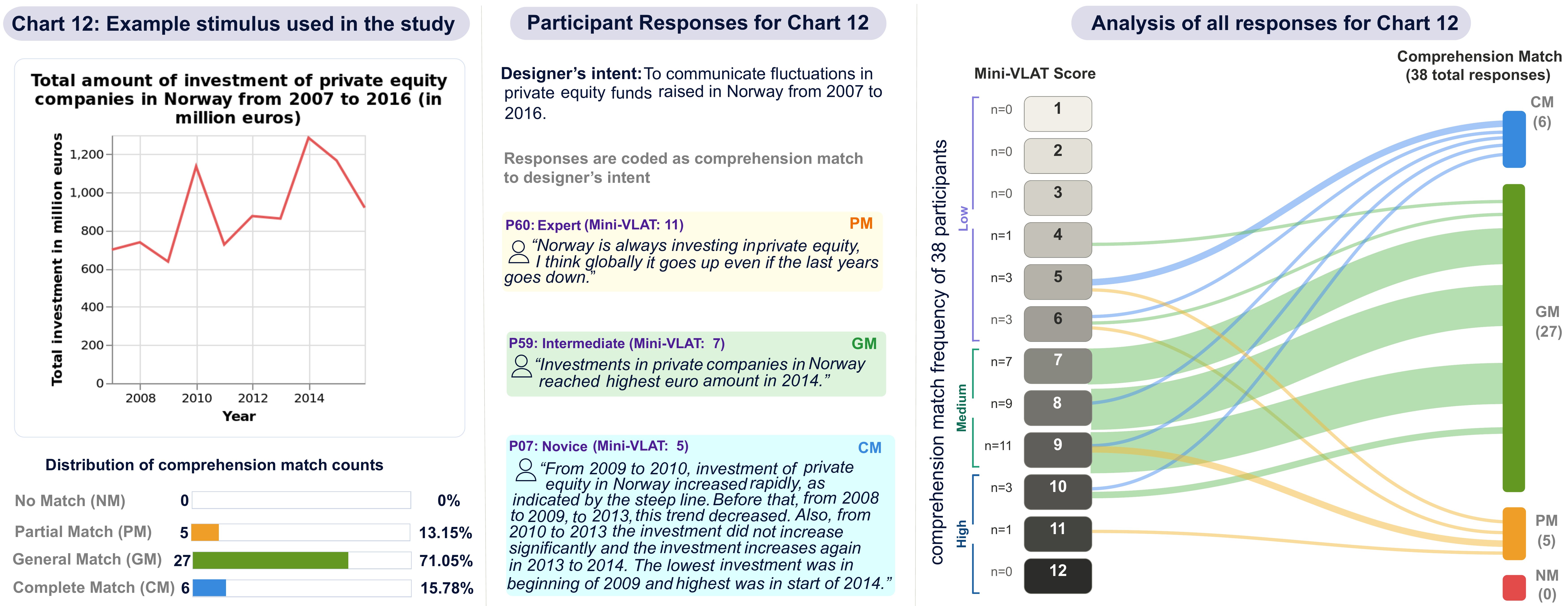}
  \caption{
  We studied how viewers with different levels of visualization literacy, measured using the Mini-VLAT~\cite{pandey2023mini}, intuitively build overall knowledge of the data without explicit cuing.  The left panel shows the example stimulus shown to participants and the resulting distribution of comprehension match; the alignment between a viewer's understanding of the chart and the designer's intent, coded as No (NM), Partial (PM), General (GM), and Complete (CM) Match across the 38 participants who viewed Chart 12. The middle panel shows the designer's intended message and participants' responses for Chart 12. The right panel shows a Sankey diagram mapping individual Mini-VLAT scores (grouped into Low, Medium, and High literacy) to the comprehension match levels. Most responses are GM. Higher Mini-VLAT scores do not lead to a more Complete Match (CM).
}
  \label{fig:teaser}
}

\abstract{Visualization literacy assessment tests such as VLAT and Mini-VLAT measure people's ability to perform isolated, low-level visual analytics tasks, such as estimating statistical quantities. However, it remains unclear whether these assessments also account for the overall understanding viewers intuitively gain about the data without explicit cuing or guidance, often referred to as high-level comprehension. In this work, we examine whether Mini-VLAT scores predict high-level comprehension, characterized as alignment between viewers’ uncued interpretations and designer intent.~We conducted a study with 80 participants, in which each participant completed the Mini-VLAT test and interpreted 16 charts selected from a set of 40 real-world visualizations.~Our qualitative and quantitative analyses show no significant relationship between participants' Mini-VLAT scores and high-level comprehension.~Our findings suggest while visualization literacy tests like Mini-VLAT capture viewers' ability to perform directed, low-level tasks, they do not assess high-level comprehension in natural, uncued settings. Future work should develop VLAT that accounts for high-level comprehension beyond estimating prespecified statistical quantities.%
\footnotetext{\hspace{-1.8em}*These authors contributed equally.}%
\footnotetext{\hspace{-1.8em}\ensuremath{\dagger}e-mail: \{aaryani, pandey, quadri\}@ou.edu}%
\footnotetext{\hspace{-1.8em}\ensuremath{\ddagger}e-mail: zeyuwang@cs.unc.edu}%
}

\begin{document}


\maketitle
\section{Introduction}
Visualization literacy is defined as people's underlying ability to read, interpret, and extract information from data as well as to construct data visualizations
~\cite{boy2014principled,lee2016vlat,beschi2025characterizingdatavisualizationliteracy, Ge2025AVEC}. To measure people's visualization literacy 
systematically, researchers developed standardized instruments such as the Visualization Literacy Assessment Test (VLAT)~\cite{lee2016vlat}. VLAT presents viewers with itemized, directed low-level analytical tasks, such as retrieving values, finding extrema, and sorting data~\cite{amar2005low}. These tests provide a validated and reusable way to quantify a viewer's visualization literacy with a single score, and a viewer can be considered an expert or novice in performing such tasks based on the scores achieved on the test~\cite{chang2025tell}. 

Building on VLAT, the Mini-VLAT~\cite{pandey2023mini} was recently introduced as a shorter, 12-item alternative to full VLAT~\cite{lee2016vlat}. Researchers have also developed assessments for specific dimensions of visualization literacy, such as CALVI, which evaluates critical reasoning about potentially misleading visualizations~\cite{ge2023calvi}. 
Existing visualization literacy assessment tests~\cite{lee2016vlat, pandey2023mini, ge2023calvi} present viewers with itemized, directed, low-level statistical tasks to measure their visualization literacy. However, understanding a graph in real-world settings involves more than estimating directed low-level statistical quantities~\cite{freedman2002toward}. Viewers in the wild intuitively build an overall understanding of a graph by eliciting key statistics, insights, and patterns, without being explicitly told which specific low-level task to perform, a process referred to as high-level comprehension~\cite{quadri2024you}. Currently, it remains unclear whether existing literacy assessment test scores also capture people's high-level comprehension in natural, uncued settings. 

To examine whether visualization literacy assessment test scores predict high-level visualization comprehension, we conducted a user study with 80 participants.~We characterized each participant's visualization literacy using the Mini-VLAT test scores and
asked them to provide natural-language interpretations of 16 charts, drawn from a pool of 40 real-world visualizations.~We considered Mini-VLAT~\cite{pandey2023mini} among other visualization literacy assessment tests to keep overall study time manageable, as it is a short-form alternative and provides results consistent with the full VLAT ($r = 0.75$).~Following prior work, we coded each interpretation into one of four comprehension-match levels: No Match, Partial Match, General Match, and Complete Match, based on its alignment with the designer’s stated objective~\cite{quadri2024you, jeon2026llms}.~We then conducted quantitative and qualitative analyses to examine whether Mini-VLAT scores predict the comprehension match outcomes and to identify patterns of high-level comprehension across Mini-VLAT scores.
%
%
If visualization literacy predicts high-level comprehension, participants with higher Mini-VLAT scores should produce interpretations that more closely match the designer's intended message.
Therefore, following the conventional characterization of viewers as experts or novices based on their literacy scores~\cite{chang2025tell}, we grouped participants by Mini-VLAT score into three levels: novices (scores 3–6), intermediate viewers (7–9), and experts (10–11). 

Our results contradict the conventional expectation that expertise based on VLAT scores leads to chart interpretations that better align with the designer's intent.~Experts should produce a higher frequency of General and Complete Match responses that align closely with the designer's intent. 
Novices should have a higher frequency of No Match and Partial Match responses, indicating interpretations that fail to align. However, in the results and analyses, we observe no such pattern.
The Mini-VLAT scores were not significantly associated with comprehension match, and our qualitative results do not show a discernible pattern between participant groups and their comprehension match outcomes. Novices' interpretations, in fact, aligned completely with designer intent more often than experts'. As shown in \autoref{fig:teaser}, a novice (Mini-VLAT 5) produced a complete match, while an expert (Mini-VLAT 11) provided only a partial match. Visualization literacy, as measured by the Mini-VLAT, thereby does not account for people's high-level comprehension.


Our findings provide preliminary evidence that 
visualization literacy assessment test scores are insufficient to represent how effectively viewers interpret a designer's intended message. The Mini-VLAT effectively captures chart-reading ability on low-level, directed tasks, while high-level comprehension additionally depends on factors such as chart context, design complexity, individual differences, domain familiarity, and the salience of specific statistics and patterns in the visualization. We, therefore, recommend that empirical studies not treat Mini-VLAT scores as a proxy for high-level comprehension skills, and that future work incorporate high-level comprehension into current assessment tests or develop dedicated measures that capture how viewers build overall knowledge from visualizations in natural, uncued settings.

\section{Related Work}
To contextualize the relationship between visualization literacy and high-level visualization comprehension, we review prior work on graphical perception, high-level visualization comprehension, visualization literacy, and the methods used to assess them.

\subsection{Graphical Perception and Comprehension}
Early graphical perception research characterized graph reading as the visual decoding of graphically encoded information and experimentally evaluated the accuracy of elementary perceptual tasks for extracting quantitative values~\cite{szafir2023visualization,cleveland84}. Subsequent research organized visual analytics into common low-level tasks, including retrieving values, filtering data, computing derived values, identifying extrema, sorting values, characterizing distributions, detecting anomalies, identifying clusters, and examining correlations~\cite{amar2005low,quadri2021survey}, and has produced rankings of how effectively different visual encodings support such tasks~\cite{davis24,mayr2011strategies}. Task-based evaluations further show that the effectiveness of basic visualization types varies across analytical tasks, thereby guiding the matching of chart types to specific analytical goals~\cite{saket2018task,behrisch2018quality}. These studies establish an empirical foundation for selecting visual encodings and chart types that enable viewers to perform specific analytical tasks accurately.

However, these studies primarily evaluate viewers’ performance on directed low-level tasks, and their findings may not fully explain the broader knowledge viewers construct from a visualization without explicit guidance~\cite{quadri2024you}.
\textit{High-level visualization comprehension} concerns how a viewer
naturally elicits the intended message from a chart, reflecting the
salient statistics and patterns that emerge organically from a
particular combination of data and design~\cite{quadri2024you}.
Recent works examine high-level comprehension through comprehension match, which captures the extent to which a viewer’s interpretation aligns with the designer’s intended message, coded as no match, partial match, general match, and complete match~\cite{quadri2024you,jeon2026llms}.
In our study, we examine whether visualization literacy, as measured by the Mini-VLAT, accounts for viewers’ high-level comprehension, using the same four-level comprehension match coding scheme.

\subsection{Visualization Literacy and Measurement}
Visualization literacy generally refers to a person’s ability to read, interpret, reason with, and extract information from visual data representations~\cite {lee2016vlat, boy2014principled, beschi2025characterizingdatavisualizationliteracy}. Prior works suggest that visualization literacy is a multidimensional construct that extends beyond basic chart reading, encompassing the competencies required to consume, create, interpret, critique, and engage with visualizations across different contexts~\cite{varona2025state, Ge2025AVEC}.
Earlier work on graph literacy also proposed broader frameworks for assessing how viewers read, interpret, and reason with graphical displays~\cite{galesic2011graph,borner2016investigating}.~Börner et al. frame data visualization literacy through definitions, conceptual frameworks, exercises, and assessments for developing and evaluating visualization literacy~\cite{borner2019survey}. Recent multiliteracy work broadens this view by arguing that visualization literacy encompasses additional competencies required for interactive, contextual, and system-level engagement with visualizations~\cite{molina2026interactiveliteracy}. These works characterize visualization literacy as a multidimensional construct that includes, but is not limited to, the ability to perform low-level tasks.

\begin{figure*}[!t]
\centering
\includegraphics[width=\textwidth]{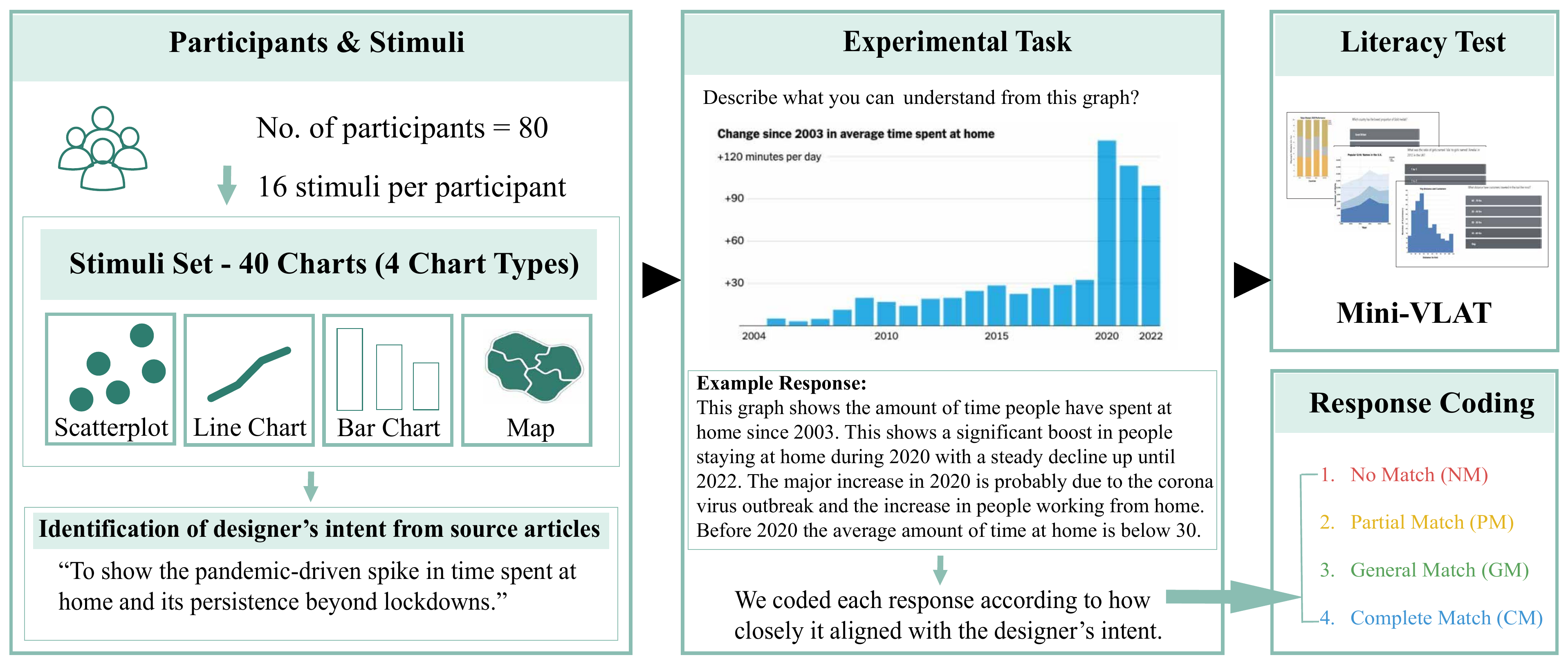}
\caption{Overview of the study design. Our study involved 80 participants, with each participant viewing 16 visualization stimuli selected from a set of 40 charts across four chart types: scatterplots, line charts, bar charts, and maps. For each chart, participants provided an open-ended description of what they understood from the graph, and their visualization literacy was measured using the Mini-VLAT test~\cite{pandey2023mini}. We identified the designer’s intent from the corresponding source articles. We coded participant responses based on their alignment with that intent into four comprehension match levels: No Match, Partial Match, General Match, and Complete Match.}

\label{fig:study_design}
\end{figure*}

Visualization Literacy Assessment Tests provide a standard approach for assessing visualization literacy through a series of directed chart-reading tasks~\cite{lee2016vlat}. Several tests, including VLAT, CALVI, adaptive visualization-literacy assessments, and Mini-VLAT, quantify visualization literacy through scores that support comparisons of chart-reading ability across users~\cite{lee2016vlat,ge2023calvi,cui2023adaptive,pandey2023mini}. Other assessment work has also examined literacy for specific visualization types, such as parallel coordinates and treemaps~\cite{firat2022p,firat2020treemap}.
Building on these works, later work expanded and refined literacy assessment. CALVI incorporated critical thinking and reasoning into visualization literacy evaluation~\cite{ge2023calvi}, while adaptive visualization literacy assessments used item difficulty and participant performance to provide more efficient and personalized measurement~\cite{cui2023adaptive}. Beyond itemized task-based tests, subjective measures such as the Subjective Graph Literacy (SGL) scale~\cite{garcia2016measuring} estimate a viewer's literacy based on self-reported performance.  More recently, Mini-VLAT~\cite{pandey2023mini} was introduced as a shorter, 12-item, timed alternative to VLAT~\cite{lee2016vlat}, evaluating visualization literacy through structured chart-reading tasks and showing results consistent with the full VLAT ($r = 0.75$). Chang et al. further used visualization-literacy scores to distinguish viewers with different literacy levels, including novice and expert groups~\cite{chang2025tell}. In our study, we use Mini-VLAT scores to characterize participants’ visualization literacy and examine whether visualization literacy, as measured using Mini-VLAT, accounts for high-level comprehension.

\section{Methodology}
\begin{figure*}[t]
  \centering
  \vspace{-4pt}
  \includegraphics[width=\textwidth]{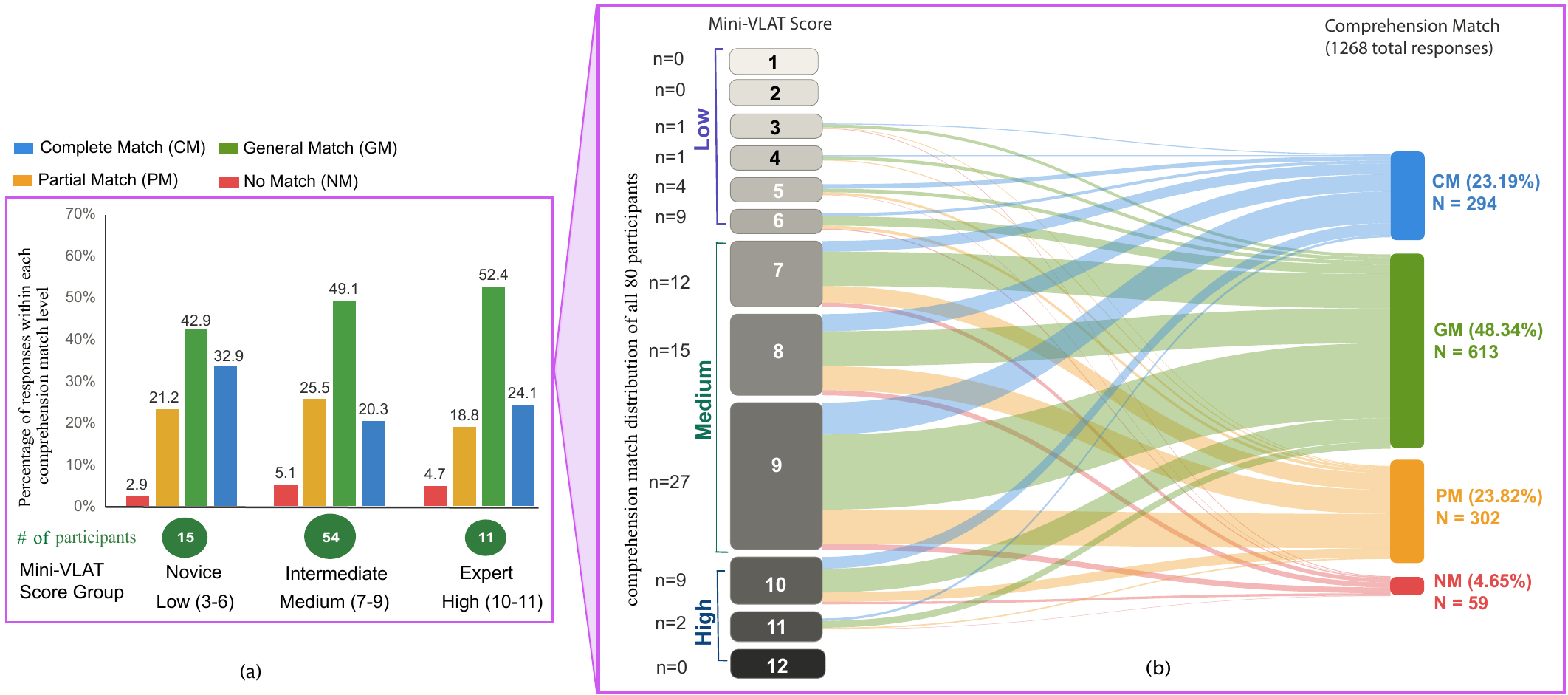}
  \caption{Distribution of comprehension match counts across participants with varying Mini-VLAT scores~\cite{pandey2023mini}. Left (a) groups participants into novice, intermediate, and expert based on their Mini-VLAT score groups, while the right panel (b) shows comprehension match counts for individual scores ranging from 1 to 12 in a Sankey diagram. In (a), bars show the percentage of responses within each match level, and the bubbles below indicate the number of participants in each group; in (b), node heights are proportional to the number of responses, the label `n' indicates the participant across each score, and `N' represents response count for each comprehension match level.}
  \vspace{-10pt}
  \label{fig:all_data}
\end{figure*}

To examine whether visualization literacy accounts for high-level visualization comprehension, we conducted a user study in which we measured participants’ visualization literacy using the Mini-VLAT and assessed their uncued interpretations of static visualizations using comprehension match coding.~\autoref{fig:study_design} shows the overall study design. 
\subsection{Stimuli Selection}

To create a diverse stimulus set, we selected 40 real-world visualizations that varied across two design factors: \textit{chart type} spanning four commonly studied types: bar charts, line charts, scatterplots, and maps, and \textit{data type}: single-class and multi-class.~We collected a total of 40 charts from various sources: 11 from ~\cite{quadri2024you}, 20 from news and analytics outlets---\textit{Statista} (6), \textit{The Economist} (4), \textit{The New York Times} (4), and one each from \textit{Reuters}, \textit{CNN}, the \textit{Financial Times}, \textit{The Wall Street Journal}, \textit{Canary Media}, and \textit{Vox}---7 from government and public data sources, such as the U.S.\ EIA, USDA, FRED, StatsAmerica, USAFacts, the Tax Foundation, and the American Communities Project, and 2 from charting-tool galleries (Google Charts and AnyChart). We also identified the designer's intent of each chart, as described in Section~\ref{sec:method}.
%

To enhance ecological validity, we selected visualizations covering five real-world topic areas: environment \& infrastructure ($13$), economy \& business ($13$), health \& well-being ($6$), society \& culture ($6$), and science \& technology ($2$).~To avoid external bias in comprehension and capture natural understanding, we removed slanted titles, statistics, and messages~\cite{kong19, kong18}.
\subsection{Participants \& Procedure}
We conducted an \textit{a priori} power analysis for linear regression, which indicated that detecting a medium effect of visualization literacy (\(f^2 = .15\), \(\alpha = .05\), \(1-\beta = .80\)) required a minimum analyzable sample of \(N = 55\). We, therefore, aimed to recruit \(N = 80\) participants to account for attrition.
We conducted a crowdsourced study with 80 participants, comprising 27 participants from the university and 53 from Prolific. We distributed flyers across the university to recruit interested participants. We recruited participants who self-reported no history of color blindness.  Participants ranged in age from 18 to 63 years and resided in English-speaking countries, including the United Kingdom, the United States, Ireland, Australia, Canada, and New Zealand.

We conducted the user study on a dedicated website, where participants signed an IRB-approved consent form. The website provides step-by-step instructions for the study, accompanied by a demonstration video. 
Each participant viewed 16 charts selected from a pool of 40 via stratified sampling, which ensured that chart and data types were equally represented within participants and balanced across the full sample. Although individual charts were not presented equally often, each chart type and data type condition was equally represented at the design-factor level.
Each participant completed two tasks: (a) describing what they understood from each graph, and (b) completing the Mini-VLAT test. We collected free-form, natural-language responses from task (a) to assess participants' understanding of each chart. We used the number of correct responses on the 12-item Mini-VLAT from task (b) as each participant's visualization literacy score.
\vspace{-2pt}
\subsection{Designer's Intent and Comprehension Match}
\label{sec:method}
 We characterize a visualization’s designer’s intent in terms of the analytical and communicative goals and messages presented in the original news articles from which the visualization was sourced, following the prior work~\cite{quadri2024you}.
We assessed comprehension match: how closely a participant’s understanding aligns with that
intended objective using the four-level coding scheme established in prior work~\cite{quadri2024you}.

\subsubsection{Designer's Intent}
%
For charts across the sources, the designer's intent is communicated through the title, caption, and surrounding text, providing a basis for inferring the intended message against which viewers' interpretations can be compared.
One primary author extracted the designer's intent for each chart by identifying the chart's title and the accompanying news article's headline, collecting the associated descriptions and captions, and merging them into a short description of the intended objective. Another author reviewed and verified the extracted intent to ensure it accurately reflected the source content, and both coders reached agreement. We treated this step as a verification process rather than an independent coding task.~\autoref{fig:teaser} shows the extracted designer's intent for an example chart (Chart 12), and the supplemental material provides the designers' intent for all 40 charts.

\subsubsection{Comprehension Match Coding}
Two authors independently coded participants' text responses into four levels of comprehension match (No Match, Partial Match, General Match, and Complete Match), defined by the extent to which participants' interpretation matched the designers' intent. The levels are defined in prior work by Quadri et al.~\cite{quadri2024you} as below :
\begin{enumerate} [label=(\roman*), nosep]
    \item Complete Match (CM): Specific statistics and patterns in the response matched the stated objective. A complete match indicates a strong alignment between participant responses and design objectives. In such cases, participants have effectively comprehended the key statistical information and insights intended by the original graph. For example, participant P07's response is a Complete Match to the designer's intent (see~\autoref{fig:teaser}).
    \item General Match (GM): Participant’s responses indicated the general knowledge they took from the graph aligned with the design’s stated objective, but not specific statistics or patterns. While there might be some minor differences or details that participants did not capture precisely, the core message and insights align well. For example, participant P59's response is a General Match to the designer's intent (see~\autoref{fig:teaser}).
    \item Partial Match (PM): Participant’s responses partially matched the stated objective. They missed the main statistical quantities or general knowledge about the data. For example, participant P60's response is a Partial Match to the designer's intent (see~\autoref{fig:teaser}).
    \item No Match (NM): Participant’s responses do not match the stated objective and missed critical information entirely. Participants have not effectively comprehended or articulated the information the graph was designed to convey. A no-match implies that participants may have misunderstood or missed intended information. 
\end{enumerate}

We calculated inter-rater reliability for comprehension match coding, resulting in substantial agreement with Cohen's $\kappa = 0.734 $.~Two coders resolved  disagreement by discussing the disputed response against the chart's designer intent until they reached a consensus label.~The coding resulted in 48.34\% ($N=613$), 23.82\% ($N=302$), 23.19\% ($N=294$), and 4.65\% ($N=59$) of the responses as GM, PM, CM, and NM, respectively (see~\autoref{fig:all_data}).

\section{Results and Analysis}
Our objective is to examine whether visualization literacy, measured using Mini-VLAT scores~\cite{pandey2023mini}, predicts people's high-level visualization comprehension patterns. 
We collected 1,280 written responses across 16 charts and 80 participants (see~\autoref{fig:study_design}). We removed 12 responses that were missing from the database resulting in 1,268 responses to analyse.
We qualitatively coded the open-ended, natural-language responses into comprehension match levels (see Section~\ref{sec:method}), then used quantitative modeling and qualitative analysis to examine whether Mini-VLAT scores were associated with comprehension match levels.
\begin{figure*}[!t]
  \centering
  \includegraphics[width=0.90\textwidth]{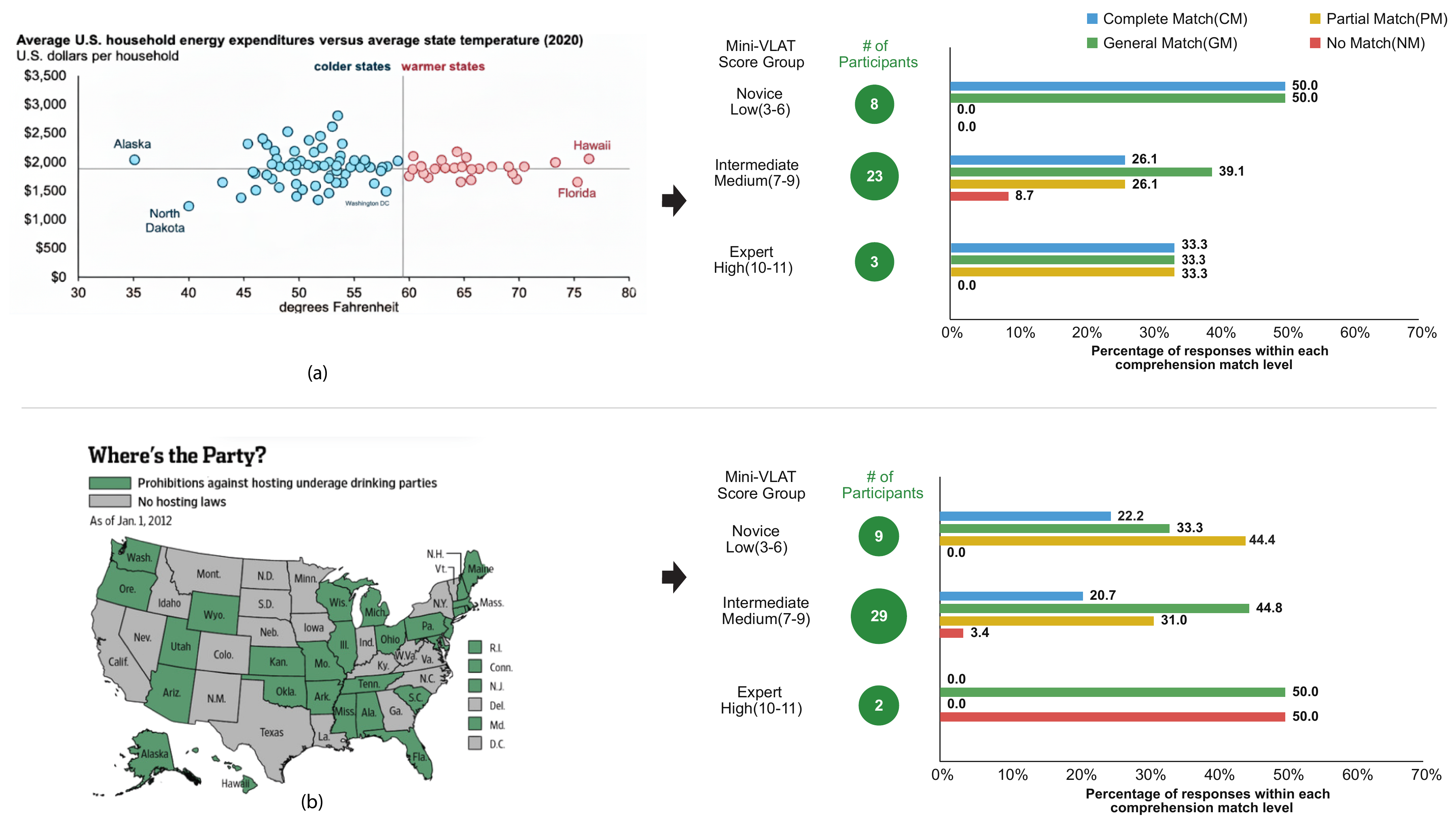}
  \caption{Example stimuli and corresponding comprehension match distributions by Mini-VLAT~\cite{pandey2023mini} score groups. Panel (a) shows a scatterplot of average U.S. household energy expenditures versus average state temperature, while panel (b) shows a map of U.S. state laws prohibiting the hosting of underage drinking parties as of January 2012. For each stimulus, the bar chart on the right summarizes how participant responses were classified across comprehension match levels: Complete Match, General Match, Partial Match, and No Match. Bars show the percentage of responses within each Mini-VLAT group, and the circles indicate the number of participants in each group.
}
  \label{fig:findings}
\end{figure*}

\subsection{Quantitative Analysis}
To test whether the Mini-VLAT scores affect the comprehension match outcomes, and thus whether visualization literacy predicts high-level comprehension, we conducted a quantitative analysis using an ordinal mixed-effects model.
We chose this model because of the ordinal nature of our comprehension match levels.

We treated the Mini-VLAT score as a fixed effect variable, while participant ID and chart ID were treated as random effects to account for individual differences and stimulus variance, with comprehension match level as the ordinal dependent variable. We fitted the model to all 1,268 responses from 80 participants across 40 charts, with each participant viewing 16 charts.
Therefore, we fit a cumulative link mixed model with a logit link where $i$ indexes participants and $j$ indexes charts:
\begin{equation}
\text{Match}_{ij} \sim \text{Score}_i + (1 \mid \text{Chart}_j) + (1 \mid \text{Participant}_i)
\label{eq:clmm}
\end{equation}
\noindent Here, $\text{Match}_{ij}$ is the comprehension match level for participant $i$ on chart $j$, and $\text{Score}_i$ is participant $i$'s Mini-VLAT score.

~\autoref{tab:qresults} shows the key results from the model, while the full results table is present in the supplementary material. Mini-VLAT score showed no significant effect on comprehension match ($estimate = -0.052$, $p = 0.58$), indicating that, after accounting for participant and chart differences, literacy score did not meaningfully explain differences in comprehension. 
The effect is small: even going from the lowest to the highest Mini-VLAT score in our data (3 to 11) shifts predicted comprehension by only about 0.42 on the model's latent scale.
By contrast, the random-effects standard deviations for participant ID ($SD = 1.234$) and chart ID ($SD = 0.792$) are substantially larger, suggesting
that high-level comprehension may be shaped more strongly by individual differences and design-specific differences than by visualization literacy alone.
\begin{table}[h]
\centering
\vspace{-4pt}
\small

\caption{Results from the cumulative link mixed model. The Mini-VLAT fixed effect is non-significant.}

\begin{tabular}{@{}llr@{}}

\toprule

\textbf{Component} & \textbf{Term} & \textbf{Value} \\

\midrule

Fixed effect   & Mini-VLAT score (estimate) & $-0.052$ \\

               & \quad $p$-value            & $0.58$  \\

\midrule

Random effect  & Participant ID (SD)        & $1.234$ \\

(Std.\ Dev.)   & Chart ID (SD)              & $0.792$ \\

\bottomrule

\end{tabular}
\label{tab:qresults}
\end{table}
\subsection{Qualitative Analysis}
To examine whether visualization literacy is reflected in high-level comprehension, we qualitatively analyzed the distribution of comprehension match responses across literacy scores (see ~\autoref{fig:all_data}).
The range of Mini-VLAT scores achieved by participants was from 3 to 11.~To examine grouped patterns, we partitioned scores into three visualization literacy groups: Low (3–6), Medium (7–9), and High (10–11), representing relatively novice, intermediate, and expert viewers, respectively~\cite{chang2025tell}.~Therefore, we expect that higher visualization literacy should lead to better high-level comprehension: the expert group should produce a higher frequency of General Match (GM) and Complete Match (CM) responses than the intermediate and novice groups.~In contrast, the novice group should produce a higher frequency of No Match (NM) and Partial Match (PM) responses than the other two.

The expected pattern of higher-literacy groups yielding more General and Complete Match responses, and fewer No and Partial Match responses did not hold across any of the comprehension match levels (see ~\autoref{fig:all_data}a).~Complete Match frequency was highest among novices at 32.9\%, decreased to 20.3\% among intermediate viewers, and then increased to 24.1\% among experts.~General Match frequency was lower among novices at 42.9\%, then increased and remained relatively stable among intermediate and expert viewers at 49.1\% and 52.4\%, respectively. Partial Match frequency increased from 21.2\% among novices to 25.5\% among intermediate viewers before decreasing to 18.8\% among experts. No Match frequency remained low across all three groups, at 2.9\% among novices, 5.1\% among intermediate viewers, and 4.7\% among experts. Comprehension match frequencies did not increase monotonically with literacy across the three groups.

The three visualization literacy groups contain unequal numbers of participants, which limits direct comparison of the group-level percentages.~We further inspected comprehension match at the level of individual Mini-VLAT scores (see ~\autoref{fig:all_data}b). Across the 1,268 responses, the overall distribution was 48.34\% GM, 23.82\% PM, 23.19\% CM, and 4.65\% NM. The distributions varied across scores without a clear trend: for instance, the score 4 group showed 0 NM responses, whereas the highest-scoring participants with score 11 still included NM response.~Complete Match responses appear across both low and high Mini-VLAT scores, but their proportions do not increase consistently with score; instead, novices have more CM than experts. General Match responses remain close across several score levels without a clear trend. These contrasts indicate that higher Mini-VLAT scores do not systematically correspond to stronger comprehension match outcomes. The grouped and individual score observations yield the same result where comprehension match outcomes did not steadily improve with Mini-VLAT score.

The chart-specific distributions of comprehension match and the literacy groups reinforce our finding that visualization literacy is not associated with alignment to the designer's intent (see~\autoref{fig:findings}).
For Chart (a), novices produced no Partial Match or No Match responses, with their responses evenly divided between Complete Match and General Match (50.0\% each), whereas experts showed an even distribution across Complete Match, General Match, and Partial Match (33.3\% each). For Chart (b), experts produced no Complete Match responses, while novices produced two Complete Match responses and no No Match responses.

These observations highlight the absence of a consistent pattern between Mini-VLAT and high-level comprehension, indicating that higher VLAT scores do not lead to interpretations that align with the designer's intent.~The distribution of comprehension matches did not consistently improve from Low (novices) to Medium (intermediates) to High (experts) Mini-VLAT score groups. Experts' responses did not always show stronger alignment with designer intent, while novices were still able to produce Complete Match responses. This supports our quantitative results: Mini-VLAT scores, although useful for characterizing visualization literacy, are insufficient to explain high-level visualization comprehension.
\label{sec-analysis}

\section{Discussion}
Our findings suggest a distinction between people's directed, low-level chart-reading ability and high-level comprehension, in which viewers naturally build knowledge from a visualization without explicit cues. We discuss two implications: that visualization literacy does not necessarily predict high-level comprehension, and that assessing high-level comprehension requires dedicated measures.

\subsection{Mini-VLAT scores Do Not Predict High-Level Comprehension}

Our analysis showed no distinguishable pattern between visualization literacy scores and frequency of comprehension matches, suggesting that itemized measures like Mini-VLAT do not fully account for whether viewers' interpretations align with designers' intents or what knowledge they intuitively build about the data.~This has a direct consequence for how literacy scores are used in visualization research: a Mini-VLAT score should not be treated as a proxy for whether a viewer will understand what a designer is trying to communicate. This disconnect reflects a difference in what visualization literacy and high-level comprehension each capture.

VLAT-based assessments effectively capture performance on directed low-level tasks, whereas high-level comprehension involves naturally building an understanding of overall statistics and patterns that aligns with the designer’s intent.~Performance on the directed tasks may therefore not translate directly into stronger alignment with the designer’s intent. A viewer can score high on literacy assessments that emphasize accurately reading values and elementary tasks yet still arrive at a misaligned comprehension.~Conversely, a viewer with lower literacy scores can achieve a comprehension that closely aligns with the designer’s intent. Accurately reading a visualization, estimating statistics, and understanding a design's communicative goals are therefore distinct skills. This means literacy scores are best used to characterize a viewer's basic chart-reading ability rather than to predict interpretive outcomes.

Our findings suggest that high-level comprehension might be shaped by an interaction among the viewer, the visualization design, and the data being interpreted, rather than by visualization literacy alone.
Some charts may communicate their intended message more clearly than others, while some charts can be more complex to interpret and less trustworthy to act upon for some viewers than for others~\cite{elhamdadi2023vistrust, quadri2024you}. The comprehension process may involve further interpretative mechanisms of inference, prior knowledge, attention to context, and individual reasoning~\cite{shah2011bar,freedman2002toward,wang2026contextualization,chang2025tell,okan2012higher} so that some viewers may rely on prior knowledge, prior attitudes, chart familiarity, or reasoning strategies when interpreting a visualization~\cite{wang2026contextualization,ancker2006design,markant2023data, quadri2024you}. These factors might explain why performance on low-level tasks, and thus Mini-VLAT scores, did not translate into high-level comprehension ability in our study.~Anticipating how an audience will interpret a visualization therefore requires considering the interaction between viewer and design, rather than relying on literacy scores alone.

\subsection{Toward Dedicated Measures of High-Level Comprehension}
%
Our findings indicate that high-level comprehension cannot be adequately characterized by Mini-VLAT scores alone, raising the need for dedicated measures that directly assess this aspect of visualization understanding.~Prior work by Öney et al.~\cite{oney2025testing} suggests that VLAT-style measures capture only part of visualization literacy, particularly chart-reading performance, while underrepresenting broader aspects of visualization understanding, such as contextual interpretation, insight extraction, design judgment, and domain knowledge.~Complementing this, our findings suggest that graph-reading literacy and high-level comprehension are distinct constructs.~Our results highlight the need for assessment instruments specifically designed for high-level comprehension
, rather than treating graph-reading literacy as a proxy for it. 

Just as the VLAT and Mini-VLAT provide validated measures of visualization literacy through directed tasks while capturing top-down interpretation, a comparable instrument is needed to assess uncued, bottom-up interpretation~\cite{quadri2024you}. Such an instrument would need to assess open-ended interpretation and to evaluate the depth and alignment of what viewers take away, along with the accuracy of individual tasks.
Alternatively, we can integrate aspects of high-level comprehension into current state-of-the-art literacy assessments, extending them beyond low-level statistical tasks so that a single instrument characterizes both what a viewer can read and what they spontaneously understand.

Participant responses further indicate that domain familiarity and contextual knowledge may shape high-level visualization comprehension. For example, one Participant (P33, literacy score = 7) responded, \textit{``I do not understand the graph. I am unsure what banking penetration is, nor do I understand how it relates to the share of respondents over time?''}. Similarly, (P52, literacy score = 10) responded \textit{``This graph is harder for me to understand; I don't do a lot in Architecture, so it's out of my range of experience''}.
These responses illustrate that participants may struggle to interpret a visualization when they lack familiarity with its domain or terminology, regardless of their Mini-VLAT scores.
Accounting for such factors would provide future research with a more complete understanding of what influences high-level comprehension, and would help identify the dimensions a dedicated measure of it should capture.
Although developing a new instrument falls outside the scope of this study, our findings highlight the need for dedicated measures of high-level comprehension as an important avenue for future work.

\section{Limitations and Future Work}
\subsection{Limitations}
We measured visualization literacy using the Mini-VLAT, where a timed format can introduce stress due to the time limit, implying th high scorers may be those who manage time pressure well rather than those who understand visualizations better~\cite{oney2025testing}. We used the Mini-VLAT rather than the full VLAT as it provided a quicker estimate of visualization literacy, so our null finding for literacy should not fully rule out a relationship that a more comprehensive assessment besides the Mini-VLAT could detect. Chart titles and framing were present in some stimuli and absent in others, as we adapted real-world visualizations and neutralized slanted titles to study uncued comprehension and capture natural understanding without external bias. This inconsistency may influence comprehension independent of viewer literacy, since the availability of a contextual title can affect how closely a response aligns with the designer's stated intent. In addition, the uneven sampling of visual literacy groups within our participant pool limits the generalizability of our results. Finally, since we recruited participants using Prolific, our samples may be limited in demographic backgrounds and may not be representative of the broader population~\cite{borgo2018crowdsourcing}, and such user studies are also subject to attrition that may bias the results.

\subsection{Future Work}
Our findings point to dedicated measures of high-level comprehension as a central direction for future work. This may take the form of a new or improved instrument for high-level comprehension, or the integration of high-level comprehension into existing literacy assessments beyond low-level statistical tasks.
Individuals' literacy could be captured by also assessing their subjective experience with visualizations, such as their self-reported familiarity with charts and the frequency with which they use and interact with graphs, which may shape high-level comprehension. Beyond developing new instruments, further evaluation is needed to determine which existing measure: full VLAT, CALVI, or subjective self-assessment most closely reflects high-level comprehension; a larger evaluation should consider these multiple measures within a single study.

Prior work also shows that visualization task performance can vary across professional domains and can relate to cognitive abilities and individual differences~\cite{hall2021professional,liu2020survey}, so future studies should sample viewers across a broader range of domains and backgrounds.
Since visualization literacy cannot be controlled during recruitment, future studies should also target balanced sampling across visualization literacy groups to reduce the effect of uneven group sizes. Finally, future work should systematically vary the presence and absence of visual elements such as titles, labels, and legends to isolate their effects.

\section{Conclusion}

We conducted a user study with 80 participants and found preliminary evidence that visualization literacy, as measured by Mini-VLAT scores, does not account for viewers’ high-level comprehension. Although visualization literacy captures important aspects of visualization understanding, particularly graph reading and the estimation of low-level statistical quantities, our findings suggest that it does not capture how viewers interpret a designer’s intended message. While low-level chart-reading abilities and high-level comprehension are related, they cannot be adequately characterized by a single assessment instrument. Future research should therefore develop and evaluate measures that directly assess high-level comprehension while accounting for factors such as chart context, design complexity, individual differences, domain familiarity, and the salience of specific statistics and patterns in visualizations.

\clearpage
\newpage
\bibliographystyle{abbrv-doi-hyperref-narrow}\bibliography{main}

@INPROCEEDINGS{amar2005low,
  author={Amar, R. and Eagan, J. and Stasko, J.},
  booktitle={IEEE Symposium on Information Visualization, 2005. INFOVIS 2005.}, 
  title={Low-level components of analytic activity in information visualization}, 
  year={2005},
  volume={},
  number={},
  pages={111-117},
  doi={10.1109/INFVIS.2005.1532136}
  }

@ARTICLE{boy2014principled,
author={Boy, Jeremy and Rensink, Ronald A. and Bertini, Enrico and Fekete, Jean-Daniel},
journal={ IEEE Transactions on Visualization \& Computer Graphics },
title={{ A Principled Way of Assessing Visualization Literacy }},
year={2014},
volume={20},
number={12},
ISSN={1941-0506},
pages={1963-1972},
doi={10.1109/TVCG.2014.2346984},
publisher={IEEE Computer Society},
address={Los Alamitos, CA, USA},
month=dec}

@inproceedings{quadri2024you, series={CHI ’24},
   title={Do You See What I See? A Qualitative Study Eliciting High-Level Visualization Comprehension},
   DOI={10.1145/3613904.3642813},
   booktitle={Proceedings of the CHI Conference on Human Factors in Computing Systems},
   publisher={ACM},
   author={Quadri, Ghulam Jilani and Wang, Arran Zeyu and Wang, Zhehao and Adorno, Jennifer and Rosen, Paul and Szafir, Danielle Albers},
   year={2024},
   month=May, pages={1–26},
   collection={CHI ’24} }

@article{jeon2026llms,
year = {2026},
author = {Jeon, Hyotaek and Lee, Hyunwook and Shin, Minjeong and Pandey, Tapendra and Kim, Joohee and Seon, Shinwook and Jeong, Daeun and Ko, Sungahn and Quadri, Ghulam Jilani},
title = {How Do LLMs See Charts? A Comparative Study on High-Level Visualization Comprehension in Humans and LLMs},
journal = {Computer Graphics Forum},
pages = {e70450},
doi = {https://doi.org/10.1111/cgf.70450}
}

@article{behrisch2018quality,
author = {Behrisch, M. and Blumenschein, M. and Kim, N. W. and Shao, L. and El-Assady, M. and Fuchs, J. and Seebacher, D. and Diehl, A. and Brandes, U. and Pfister, H. and Schreck, T. and Weiskopf, D. and Keim, D. A.},
title = {Quality Metrics for Information Visualization},
journal = {Computer Graphics Forum},
volume = {37},
number = {3},
pages = {625-662},
doi = {https://doi.org/10.1111/cgf.13446},
year = {2018}
}

@ARTICLE{quadri2021survey,
  author={Quadri, Ghulam Jilani and Rosen, Paul},
  journal={IEEE Transactions on Visualization and Computer Graphics}, 
  title={A Survey of Perception-Based Visualization Studies by Task}, 
  year={2022},
  volume={28},
  number={12},
  pages={5026-5048},
  doi={10.1109/TVCG.2021.3098240}}

@ARTICLE{lee2016vlat,
  author={Lee, Sukwon and Kim, Sung-Hee and Kwon, Bum Chul},
  journal={IEEE Transactions on Visualization and Computer Graphics}, 
  title={VLAT: Development of a Visualization Literacy Assessment Test}, 
  year={2017},
  volume={23},
  number={1},
  pages={551-560},
  doi={10.1109/TVCG.2016.2598920}}

@ARTICLE{cui2023adaptive,
  author={Cui, Yuan and Ge, Lily W. and Ding, Yiren and Yang, Fumeng and Harrison, Lane and Kay, Matthew},
  journal={IEEE Transactions on Visualization and Computer Graphics}, 
  title={Adaptive Assessment of Visualization Literacy}, 
  year={2024},
  volume={30},
  number={1},
  pages={628-637},
  doi={10.1109/TVCG.2023.3327165}}

@inproceedings{ge2023calvi,
author = {Ge, Lily W. and Cui, Yuan and Kay, Matthew},
title = {CALVI: Critical Thinking Assessment for Literacy in Visualizations},
year = {2023},
isbn = {9781450394215},
publisher = {Association for Computing Machinery},
address = {New York, NY, USA},
url = {https://doi.org/10.1145/3544548.3581406},
doi = {10.1145/3544548.3581406},
booktitle = {Proceedings of the 2023 CHI Conference on Human Factors in Computing Systems},
articleno = {815},
numpages = {18},
location = {Hamburg, Germany},
series = {CHI '23}
}

@article{pandey2023mini,
   title={Mini‐VLAT: A Short and Effective Measure of Visualization Literacy},
   volume={42},
   ISSN={1467-8659},
   DOI={10.1111/cgf.14809},
   number={3},
   journal={Computer Graphics Forum},
   publisher={Wiley},
   author={Pandey, Saugat and Ottley, Alvitta},
   year={2023},
   month=jun, pages={1–11} }

@article{garcia2016measuring,
author = {Garcia-Retamero, Rocio and Cokely, Edward and Ghazal, Saima and Joeris, Alexander},
year = {2016},
month = {07},
pages = {854-867},
title = {Measuring Graph Literacy without a Test: A Brief Subjective Assessment},
volume = {36},
journal = {Medical Decision Making},
doi = {10.1177/0272989X16655334}
}

@article{borner2019survey,
author = {Katy Börner  and Andreas Bueckle  and Michael Ginda },
title = {Data visualization literacy: Definitions, conceptual frameworks, exercises, and assessments},
journal = {Proceedings of the National Academy of Sciences},
volume = {116},
number = {6},
pages = {1857-1864},
year = {2019},
doi = {10.1073/pnas.1807180116},
}

@inproceedings{oney2025testing,
  author = {Öney, Seyda and Abdelaal, Moataz and Kurzhals, Kuno and Betz, Paul and Kropp, Cordula and Weiskopf, Daniel},
  booktitle = {2024 IEEE Evaluation and Beyond - Methodological Approaches for Visualization (BELIV)},
  doi = {10.1109/beliv64461.2024.00017},
  orcid-numbers = {Öney, Seyda/0000-0002-5785-6788, Abdelaal, Moataz/0000-0003-4630-7916, Kurzhals, Kuno/0000-0003-4919-4582, Weiskopf, Daniel/0000-0003-1174-1026},
  pages = {106-114},
  publisher = {IEEE},
  title = {Testing the Test: Observations When Assessing Visualization Literacy of Domain Experts},
  year = 2025
}

@ARTICLE{chang2025tell,
  author={Chang, Minsuk and Wang, Yao and Wang, Huichen Will and Zhou, Yuanhong and Bulling, Andreas and Bearfield, Cindy Xiong},
  journal={IEEE Transactions on Visualization and Computer Graphics}, 
  title={Tell Me Without Telling Me: Two-Way Prediction of Visualization Literacy and Visual Attention}, 
  year={2026},
  volume={32},
  number={1},
  pages={1405-1415},
  doi={10.1109/TVCG.2025.3634815}}

@inproceedings{kong19,
author = {Kong, Ha-Kyung and Liu, Zhicheng and Karahalios, Karrie},
year = {2019},
month = {04},
pages = {1-13},
title = {Trust and Recall of Information across Varying Degrees of Title-Visualization Misalignment},
isbn = {978-1-4503-5970-2},
booktitle = {CHI '19: Proceedings of the 2019 CHI Conference on Human Factors in Computing Systems},
doi = {10.1145/3290605.3300576}
}

@inproceedings{kong18,
author = {Kong, Ha-Kyung and Liu, Zhicheng and Karahalios, Karrie},
title = {Frames and Slants in Titles of Visualizations on Controversial Topics},
year = {2018},
isbn = {9781450356206},
publisher = {Association for Computing Machinery},
address = {New York, NY, USA},
url = {https://doi.org/10.1145/3173574.3174012},
doi = {10.1145/3173574.3174012},
booktitle = {Proceedings of the 2018 CHI Conference on Human Factors in Computing Systems},
pages = {1–12},
numpages = {12},
location = {Montreal QC, Canada},
series = {CHI '18}
}

@inproceedings{molina2026interactiveliteracy,
   title={A Multiliteracy Model for Interactive Visualization Literacy: Definitions, Literacies, and Steps for Future Research},
   DOI={10.1145/3772318.3793423},
   booktitle={Proceedings of the 2026 CHI Conference on Human Factors in Computing Systems},
   publisher={ACM},
   author={Molina León, Gabriela and Bach, Benjamin and Valentim, Matheus and Elmqvist, Niklas},
   year={2026},
   month=Apr, pages={1–16},
   collection={CHI ’26} }

@article{cleveland84,
  author    = {William S. Cleveland and Robert McGill},
  title     = {Graphical perception: Theory, experimentation, and application to the development of graphical methods},
  journal   = {Journal of the American Statistical Association},
  year      = {1984},
  volume    = {79},
  number    = {387},
  pages     = {531--554},
  doi       = {10.1080/01621459.1984.10478080}
}

@article{davis24,
   title={The Risks of Ranking: Revisiting Graphical Perception to Model Individual Differences in Visualization Performance},
   volume={30},
   ISSN={2160-9306},
   DOI={10.1109/tvcg.2022.3226463},
   number={3},
   journal={IEEE Transactions on Visualization and Computer Graphics},
   publisher={Institute of Electrical and Electronics Engineers (IEEE)},
   author={Davis, Russell and Pu, Xiaoying and Ding, Yiren and Hall, Brian D. and Bonilla, Karen and Feng, Mi and Kay, Matthew and Harrison, Lane},
   year={2024},
   month=Mar, pages={1756–1771} }

@article{mayr2011strategies,
author = {Eva Mayr and Michael Smuc and Hanna Risku},
title ={Many roads lead to Rome: Mapping users’ problem-solving strategies},

journal = {Information Visualization},
volume = {10},
number = {3},
pages = {232-247},
year = {2011},
doi = {10.1177/1473871611415987},
}

@article{borgo2018crowdsourcing,
author = {Borgo, R. and Micallef, L. and Bach, B. and McGee, F. and Lee, B.},
title = {Information Visualization Evaluation Using Crowdsourcing},
journal = {Computer Graphics Forum},
volume = {37},
number = {3},
pages = {573-595},
doi = {https://doi.org/10.1111/cgf.13444},

year = {2018}
}

@misc{varona2025state,
author = {Varona, Matthew and Bonilla, Karen and Hedayati, Maryam and Joshi, Alark and Harrison, Lane and Kay, Matthew and Nobre, Carolina},
year = {2025},
month = {08},
pages = {},
title = {The State of the Art in Visualization Literacy},
doi = {10.48550/arXiv.2509.01018}
}

@InProceedings{freedman2002toward,
author="Freedman, Eric G.
and Shah, Priti",
editor="Hegarty, Mary
and Meyer, Bernd
and Narayanan, N. Hari",
title="Toward a Model of Knowledge-Based Graph Comprehension",
booktitle="Diagrammatic Representation and Inference",
year="2002",
publisher="Springer Berlin Heidelberg",
address="Berlin, Heidelberg",
pages="18--30",
isbn="978-3-540-46037-4",
doi="https://doi.org/10.1007/3-540-46037-3_3"
}

@article{galesic2011graph,
  title={Graph literacy: A cross-cultural comparison},
  author={Galesic, Mirta and Garcia-Retamero, Rocio},
  journal={Medical decision making},
  volume={31},
  number={3},
  pages={444--457},
  year={2011},
  doi = {10.1177/0272989X10373805},
  publisher={Sage Publications Sage CA: Los Angeles, CA}
}

@article{borner2016investigating,
author = {Börner, Katy and Maltese, Adam and Balliet, Russell and Heimlich, J.E.},
year = {2015},
month = {08},
pages = {198-213},
title = {Investigating aspects of data visualization literacy using 20 information visualizations and 273 science museum visitors},
volume = {15},
journal = {Information Visualization},
doi = {10.1177/1473871615594652}
}

@article{firat2022p,
author = {Firat, Elif E. and Denisova, Alena and Wilson, Max and Laramee, Robert},
year = {2022},
month = {06},
pages = {},
title = {P-Lite: A study of parallel coordinate plot literacy},
volume = {6},
journal = {Visual Informatics},
doi = {10.1016/j.visinf.2022.05.002}
}

@inproceedings{firat2020treemap,
booktitle = {Eurographics 2020 - Education Papers},
editor = {Romero, Mario and Sousa Santos, Beatrice},
title = {{Treemap Literacy: A Classroom-Based Investigation}},
author = {Firat, Elif E. and Denisova, Alena and Laramee, Robert S.},
year = {2020},
publisher = {The Eurographics Association},
ISSN = {1017-4656},
ISBN = {978-3-03868-102-1},
DOI = {10.2312/eged.20201032}
}

@article{shah2011bar,
author = {Shah, Priti and Freedman, Eric G.},
title = {Bar and Line Graph Comprehension: An Interaction of Top-Down and Bottom-Up Processes},
journal = {Topics in Cognitive Science},
volume = {3},
number = {3},
pages = {560-578},
doi = {https://doi.org/10.1111/j.1756-8765.2009.01066.x},
year = {2011}
}

@book{szafir2023visualization,
  title={Visualization psychology},
  author={Szafir, Danielle Albers and Borgo, Rita and Chen, Min and Edwards, Darren J and Fisher, Brian and Padilla, Lace},
  year={2023},
  publisher={Springer},
  doi="10.1007/978-3-031-34738-2"
}

@article{wang2026contextualization,
   title={Contextualization or Rationalization?
                    The Effect of Causal Priors on Data Visualization Interpretation},
   volume={32},
   ISSN={2160-9306},
   DOI={10.1109/tvcg.2026.3663050},
   number={7},
   journal={IEEE Transactions on Visualization and Computer Graphics},
   publisher={Institute of Electrical and Electronics Engineers (IEEE)},
   author={Wang, Arran Zeyu and Borland, David and Calcaterra, Estella and Gotz, David},
   year={2026},
   month=jul, pages={5113–5127} }

@ARTICLE{beschi2025characterizingdatavisualizationliteracy,
  author={Beschi, Sara and Falessi, Davide and Golia, Silvia and Locoro, Angela},
  journal={IEEE Access}, 
  title={Characterizing Data Visualization Literacy for Standardization: A Systematic Literature Review}, 
  year={2025},
  volume={13},
  number={},
  pages={65704-65725},
  doi={10.1109/ACCESS.2025.3559298}}

@article{ancker2006design,
author = {Ancker, Jessica and Senathirajah, Yalini and Kukafka, R. and Starren, J.},
year = {2006},
month = {11},
pages = {608-618},
title = {Design Features of Graphs in Health Risk Communication: A Systematic Review},
volume = {13},
journal = {Journal of the American Medical Informatics Association},
doi = {10.1197/jamia.M2115}
}

@article{okan2012higher,
  title={When higher bars are not larger quantities: On individual differences in the use of spatial information in graph comprehension},
  author={Okan, Yasmina and Garcia-Retamero, Rocio and Galesic, Mirta and Cokely, Edward T},
  journal={Spatial Cognition \& Computation},
  volume={12},
  number={2-3},
  pages={195--218},
  year={2012},
  publisher={Taylor \& Francis},
doi="https://doi.org/10.1080/13875868.2012.659302"
}

@ARTICLE{elhamdadi2023vistrust,
author={Elhamdadi, Hamza and Stefkovics, Adam and Beyer, Johanna and Moerth, Eric and Pfister, Hanspeter and Bearfield, Cindy Xiong and Nobre, Carolina},
journal={ IEEE Transactions on Visualization \& Computer Graphics },
title={{ Vistrust: a Multidimensional Framework and Empirical Study of Trust in Data Visualizations }},
year={2024},
volume={30},
number={01},
ISSN={1941-0506},
pages={348-358},
doi={10.1109/TVCG.2023.3326579},
url = {https://doi.ieeecomputersociety.org/10.1109/TVCG.2023.3326579},
publisher={IEEE Computer Society},
address={Los Alamitos, CA, USA},
month=jan}

@ARTICLE{hall2021professional,
  author={Hall, Kyle Wm. and Kouroupis, Anthony and Bezerianos, Anastasia and Szafir, Danielle Albers and Collins, Christopher},
  journal={IEEE Transactions on Visualization and Computer Graphics}, 
  title={Professional Differences: A Comparative Study of Visualization Task Performance and Spatial Ability Across Disciplines}, 
  year={2022},
  volume={28},
  number={1},
  pages={654-664},
  doi={10.1109/TVCG.2021.3114805}}

@inproceedings{markant2023data,
author = {Markant, Douglas and Rogha, Milad and Karduni, Alireza and Wesslen, Ryan and Dou, Wenwen},
title = {When do data visualizations persuade? The impact of prior attitudes on learning about correlations from scatterplot visualizations},
year = {2023},
isbn = {9781450394215},
publisher = {Association for Computing Machinery},
address = {New York, NY, USA},
url = {https://doi.org/10.1145/3544548.3581330},
doi = {10.1145/3544548.3581330},
booktitle = {Proceedings of the 2023 CHI Conference on Human Factors in Computing Systems},
articleno = {174},
numpages = {16},
location = {Hamburg, Germany},
series = {CHI '23}
}

@ARTICLE{saket2018task,
  author={Saket, Bahador and Endert, Alex and Demiralp, {\c{C}}a{\u{g}}atay},
  journal={IEEE Transactions on Visualization and Computer Graphics}, 
  title={Task-Based Effectiveness of Basic Visualizations}, 
  year={2019},
  volume={25},
  number={7},
  pages={2505-2512},
  doi={10.1109/TVCG.2018.2829750}}

@article{liu2020survey,
author = {Liu, Zhengliang and Crouser, R. Jordan and Ottley, Alvitta},
title = {Survey on Individual Differences in Visualization},
journal = {Computer Graphics Forum},
volume = {39},
number = {3},
pages = {693-712},
doi = {https://doi.org/10.1111/cgf.14033},
year = {2020}
}

@inproceedings{Ge2025AVEC,
booktitle = {Proceedings of the 2025 CHI Conference on Human Factors in Computing Systems},
author = {Ge, Lily W. and Cui, Yuan and Kay, Matthew},
title = {AVEC: An Assessment of Visual Encoding Ability in Visualization Construction},
year = {2025},
isbn = {9798400713941},
publisher = {Association for Computing Machinery},
address = {New York, NY, USA},
doi = {10.1145/3706598.3713364},
articleno = {1166},
numpages = {16},
location = {
},
series = {CHI '25}
}

\end{document}